\documentclass[aps,showpacs,amsmath,amssymb,twocolumn,pra,superscriptaddress,notitlepage]{revtex4-2}

\usepackage{qcircuit}
\usepackage{amsmath,bm}
\usepackage[dvips]{graphicx}
\usepackage{amsmath,amssymb,amsthm,mathrsfs,amsfonts,dsfont}
\usepackage{subfigure, epsfig}
\usepackage{braket}
\usepackage{bm}
\usepackage{enumerate}
\usepackage{color}
\usepackage{graphicx}
\usepackage{algorithm}
\usepackage{algorithmic}
\usepackage{braket}
\usepackage{comment}
\usepackage{appendix}
\usepackage{here}
\usepackage{tabularx}

\begin{document}

% -------------------- TITLE --------------------

\title{Quantum algorithm for calculation of transition amplitudes in hybrid tensor networks}

% ------------ AUTHORS AND AFFILIATIONS ----------

\author{Shu Kanno}
\email{kanno.s.ac@m.titech.ac.jp}
\affiliation{Department of Materials Science and Engineering, Tokyo Institute of Technology, 2-12-1 O-okayama, Meguro-ku, Tokyo 152-8552, Japan}

\author{Suguru Endo}
\email{suguru.endou.uc@hco.ntt.co.jp}
\affiliation{NTT Secure Platform Laboratories, NTT Corporation, Musashino 180-8585, Japan}
\affiliation{JST, PRESTO, 4-1-8 Honcho, Kawaguchi, Saitama, 332-0012, Japan}

\author{Yasunari Suzuki}
\affiliation{NTT Secure Platform Laboratories, NTT Corporation, Musashino 180-8585, Japan}
\affiliation{JST, PRESTO, 4-1-8 Honcho, Kawaguchi, Saitama, 332-0012, Japan}

\author{Yuuki Tokunaga}
\affiliation{NTT Secure Platform Laboratories, NTT Corporation, Musashino 180-8585, Japan}

% -------------------- ABSTRACT --------------------

\begin{abstract}
The hybrid tensor network approach allows us to perform calculations on systems larger than the scale of a quantum computer. However, when calculating transition amplitudes, there is a problem that the number of terms to be measured increases exponentially with that of the contracted operators. The problem is caused by the fact that the contracted operators are represented as non-Hermitian operators. In this study, we propose a method for the hybrid tensor network calculation that contracts non-Hermitian operators without an exponential increase in the number of terms. In the proposed method, transition amplitudes are calculated by combining the singular value decomposition of the contracted non-Hermitian operators with a Hadamard test. The method significantly extends the applicability of the hybrid tensor network approach.
\end{abstract}

\maketitle

\section{introduction}
\label{introduction}
Quantum computers are expected to be capable of executing classically intractable calculations~\cite{Bharti2021-ov,Cerezo2020-un,Cao2019-qa,Endo2021-ku,Shikano2020-fm,McArdle2020-fz,Bauer2020-ug,Moll2018-xx}. It has been reported that quantum computers can outperform classical computers in some tasks~\cite{Arute2019-hj,Zhong2020-bf,Wu2021-ky}. However, quantum resource limitations are obstacles to the practical application of quantum computers. Current quantum computers are so-called noisy intermediate-scale quantum (NISQ) devices~\cite{Preskill2018-sc}, and we can control only tens to hundreds of noisy qubits on them~\cite{Peruzzo2014-kp,Kandala2017-lh,Temme2017-vo,Endo2021-ku,Endo2018-zg,Kandala2019-ze,Takagi2021-ne}. Their hardware limitation makes it difficult to apply quantum computers to practical tasks that require large numbers of qubits or deep quantum circuits~\cite{Yuan2021-ih,Fujii2020-xc,Mizuta2021-pw,Takeshita2020-ec,Yamazaki2018-pu,Bauman2019-me,Kottmann2021-hq,Huggins2019-qa,Cong2019-qb,Kim2017-ak,Liu2019-vp,Foss-Feig2020-si,Eddins2021-vn}.

\begin{figure}
\includegraphics[width=1\columnwidth]{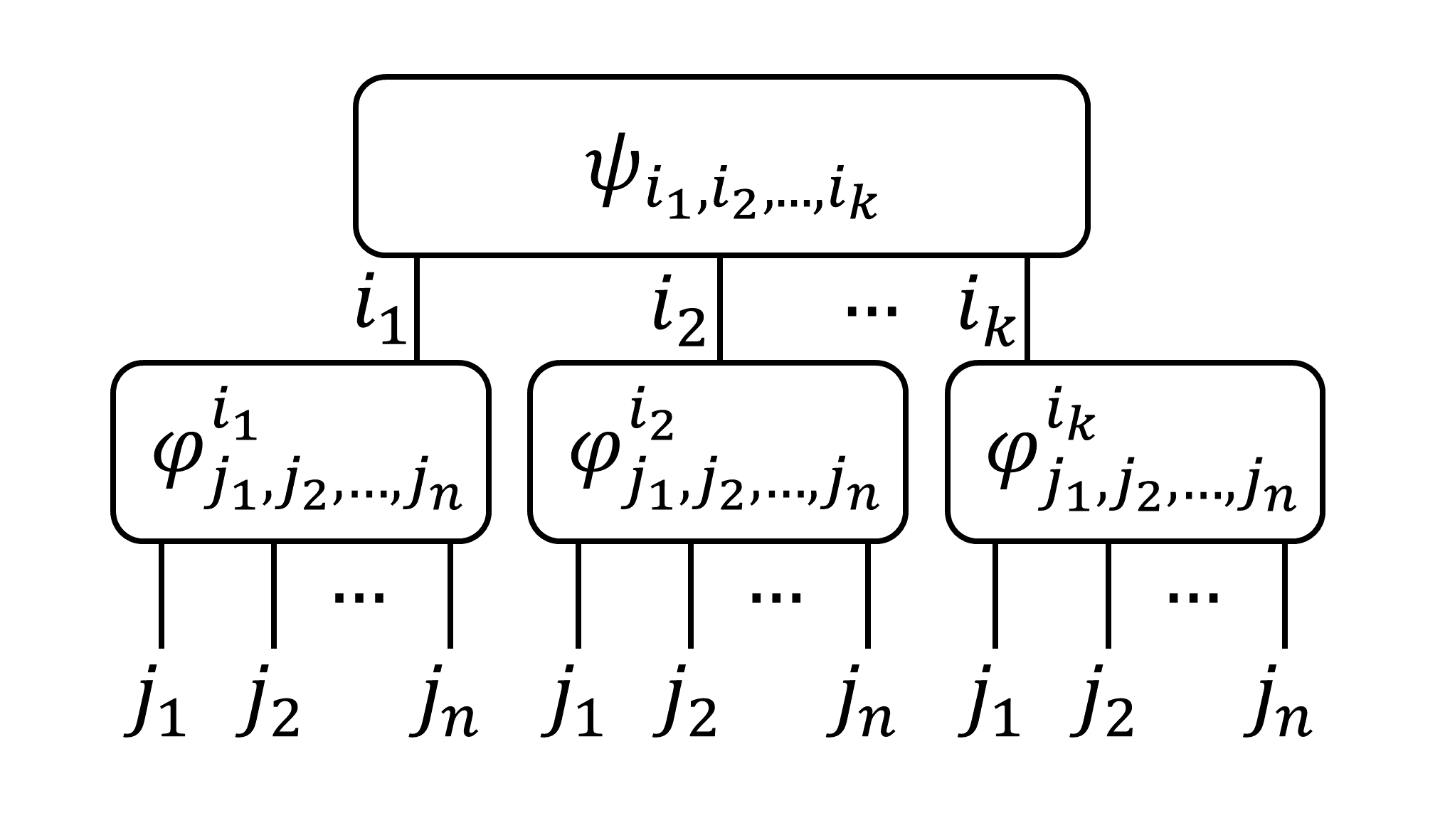}
\caption{A quantum-quantum tree-tensor network as in Eq.~(\ref{Eq:tensornetwork}). $\varphi_{j_1, j_2, \dots ,j_n}^{i_m}$ is defined as $\braket{j_1 j_2 \dots j_n | \varphi^{i_m}}$. In a quantum-classical tree-tensor network, $\psi_{i_1, i_2, \dots ,i_k}$ is replaced with a classical tensor, while in a classical-quantum tree-tensor network, classical tensors are used in place of $\varphi_{j_1, j_2, \dots ,j_n}^{i_m}$.}
 \label{fig:qqtensor}
\end{figure}

The hybrid tensor network approach has recently been proposed to overcome the limitations of NISQ devices ~\cite{Yuan2021-ih}. The approach enables the treatment of a larger number of quantum states than the number of qubits of actual quantum devices by representing quantum states as a combination of conventional classical tensors and quantum tensors. A quantum tensor has upper and lower indices, for example, $\psi_{i_1, i_2 \dots i_K}^{j_1, j_2 \dots j_L}$, which represents $K$-qubit systems indexed by an $L$-bit string. In other words, the quantum state is defined in terms of $L$ classical bits $(j_1, j_2, \dots, j_L)$ as
\begin{align}
\ket{\psi^{j_1, j_2, \dots j_L}}= \sum_{i_1, i_2, \dots, i_K} \psi_{i_1, i_2, \dots, i_K}^{j_1, j_2, \dots, j_L} \ket{i_1 i_2 \dots i_K},
\end{align}
where $\psi_{i_1, i_2, \dots, i_K}^{j_1, j_2, \dots, j_L} \in \mathbb{C}$, $\sum_{i_1, i_2, \dots, i_K} |\psi_{i_1, i_2, \dots, i_K}^{j_1, j_2, \dots, j_L}|^2 = 1$, and $\ket{i_1 i_2 \dots i_K}$ is a computational basis of the $K$-qubit Hilbert space. One of the most vital forms of the hybrid tensor network is the hybrid tree-tensor network, where a network of quantum and classical tensors composes a tree graph. While contraction of general hybrid tensor networks can be costly, hybrid tree-tensor networks can be contracted efficiently to obtain expectation values~\cite{Yuan2021-ih}. 
In this paper, we mainly discuss a two-layer hybrid tree-tensor network with only quantum tensors, which is called a quantum-quantum tree-tensor network, since it captures the essential properties of hybrid tree-tensor networks and the use of classical tensors restricts the range of representation to classically tractable quantum states such as matrix product states~\cite{Schollwock2011-im}.
A quantum-quantum tree-tensor network that expresses $k$ subsystems of $n$-qubit states is represented as

\begin{align}
\ket{\psi_{HT}}=\sum_{i_1, i_2, \dots ,i_k} \psi_{i_1, i_2, \dots ,i_k} \ket{\varphi^{i_1}} \otimes \ket{\varphi^{i_2}} \otimes \dots \otimes \ket{\varphi^{i_k}},
\label{Eq:tensornetwork}
\end{align}
where $\ket{\psi_{HT}}$ is the unnormalized state, $\psi_{i_1, i_2, \dots, i_k} = \braket{i_1 i_2 \dots i_k | \psi}$ is the probability amplitude of a $k$-qubit state $\ket{\psi}$, and $\ket{\varphi^{i_m}} (m=1, 2, \dots ,k)$ is an $n$-qubit state of the $m$-th subsystem indexed by a classical bit $i_m$. Figure~\ref{fig:qqtensor} shows the network diagram of $\ket{\psi_{HT}}$. While the state $\ket{\psi_{HT}}$ represents a $kn$-qubit state, we can efficiently calculate the expectation value of an observable $O=\bigotimes_{m=1}^k O_m$, where $O_m$ operates on $\ket{\varphi^{i_m}}$ via proper tensor contractions, using a quantum computer with $O(\max (k,n))$ qubits. The contraction for evaluating the expectation value $\braket{O}=\bra{\psi_{HT}} O \ket{\psi_{HT}}$ (without normalization) can be implemented as follows. First, we measure $O_m$ with multiple states and construct operators $M_m^{i'_m i_m} = \bra{\varphi^{i'_m}} O_m \ket{\varphi^{i_m}}$. Note that each $M_m$ is a $2 \times 2$ Hermitian matrix when $O_m$ is Hermitian. Then, since $M_m$ is Hermitian, we can measure the expectation value of $\bigotimes_{m=1}^k M_m$ for the state $\ket{\psi}$, which is equal to $\braket{O}$.

The approach allows for simulations beyond the scale of the quantum hardware. For example, the energy and spin-spin correlation functions of electrons can be calculated with this approach. However, the approach has a serious problem preventing it from being used in a range of applications: it can only calculate the expectation value of observables. In other words, there is a problem in the approach when the contracted operator $M_m$ is non-Hermitian. Hereinafter, we denote the Hermitian and non-Hermitian contracted operators as $M_m$ and $N_m$, respectively. The reason for the problem is that the number of terms to be measured increases exponentially with $k$ when $\bigotimes_{m=1}^k N_m$ is calculated naively. Specifically, non-Hermitian operators are decomposed into sums of Pauli operators, as in $\bigotimes_{m=1}^k N_m = \bigotimes_{m=1}^k (h_{Im} I_m + h_{Xm} X_m + h_{Ym} Y_m + h_{Zm} Z_m )$, where $I_m$ is the identity operator, $X_m$, $Y_m$, and $Z_m$ are Pauli operators which act on the $m$-th qubit, and $h_\alpha (\alpha \in \{ I_m, X_m, Y_m, Z_m \})$ are the corresponding coefficients, with up to $4^k$ terms appearing. One example where the problem occurs is in the calculation of the transition amplitude related to Green's functions and photon emissions/absorptions~\cite{Endo2020-ct,Ibe2020-jp}. Overlap of two quantum states, which is exploited in subroutines in many algorithms~\cite{Romero2017-kf,LaRose2019-nl,Jones2019-la,Higgott2019-az}, is a special case of the transition amplitude. Thus, the difficulty of computing the expectation value of a non-Hermitian operator limits the applicability of the hybrid tensor network approach.

In this study, we propose a method for calculating transition amplitudes with the hybrid tensor network approach. The main point of the method is the treatment of tensor products of non-Hermitian operators. In a naive calculation, an exponential number of terms in $\bigotimes_{m=1}^k N_m$ will appear. We propose two ways to avoid the problem. One is a Monte-Carlo contraction method and the other is a singular value decomposition (SVD) contraction method, and the second method is the main proposal in this paper. Although the first method can avoid having to measure all the terms whose number increases exponentially with $k$, the second method is exponentially more efficient than the first one in terms of the sampling cost.

In the following, we give an overview of quantum-quantum tree-tensor networks~\cite{Yuan2021-ih} for obtaining the expectation values of observables in Sec.~\ref{Overview of hybrid tensor network}. Then, the method of calculating transition amplitudes and overlaps is described in Sec.~\ref{Calculation of transition amplitudes and overlaps}; the contraction of quantum tensors in subsystems is discussed in Sec.~\ref{Contraction of quantum tensors in subsystems} and the contraction of non-Hermitian matrices in Sec.~\ref{Contraction of non-Hermitian matrices}. We discuss the future applications of the method in Sec.~\ref{conclusion}. 

\section{Overview of the hybrid tensor network}
\label{Overview of hybrid tensor network}
We present an overview of the hybrid tensor network simulation on the state described by Eq.~(\ref{Eq:tensornetwork})  introduced in Ref. \cite{Yuan2021-ih}. Letting the observable $O$ be $O=\bigotimes_{m=1}^k O_m$ and $O_m=\bigotimes_{r=1}^n O_{mr}$ ($r = 1, 2, \dots , n$), the expectation value of the observable including the normalization constant can be described as 
\begin{equation}
\begin{aligned}
\braket{O}&=\frac{1}{A^2} \bra{\psi_{HT}} O \ket{\psi_{HT}} \\
&=\frac{1}{A^2} \sum_{\vec{i}'~\vec{i}} \psi_{\vec{i}'}^* \psi_{\vec{i}} \prod_{m=1}^k M_m^{i_m' i_m}
\label{Eq:tensornetwork2}
\end{aligned}
\end{equation}
and
\begin{equation}
\begin{aligned}
A &= \sqrt{\braket{\psi_{HT} | \psi_{HT}}}\\
 &= \sqrt{\sum_{\vec{i}'~\vec{i}} \psi_{\vec{i}'}^* \psi_{\vec{i}} \prod_{m=1}^k M_{Am}^{i_m' i_m}},
\label{Eq:tensornetwork3}
\end{aligned}
\end{equation}
where $A$ is a normalization constant, $\vec{i}= (i_1, i_2, \dots , i_k)$ with $i_m$ taking either $0$ or $1$, $M_m^{i_m' i_m}= \bra{\varphi^{i_m'}} O_m \ket{\varphi^{i_m}}$, and $M_{Am}^{i_m' i_m}= \braket{\varphi^{i_m'} | \varphi^{i_m}}$. $M_m^{i_m' i_m}$ ($M_{Am}^{i_m' i_m}$) is an element of the $2 \times 2$ matrix $M_m$ ($M_{Am}$). When $O_m$ is assumed to be an observable, i.e., a Hermitian operator, $M_m$ also becomes a Hermitian operator. $M_{Am}$ is a Hermitian operator since $M_{Am}$ is a special case of $O_m = I_m$ in $M_m$, where $I_m$ is the identity operator.

The procedure for constructing $M_m$ and $M_{Am}$ depends on how the indices $i_m$ of the wave function $\ket{\varphi^{i_m}}$ are mapped. We suppose there are two cases of the mapping; one case is where the indices $i_m$ are mapped to unitary gates, i.e., $\ket{\varphi^{i_m}}=U_{Cm}^{i_m} \ket{0}^{\otimes n}$; the other case is where the indices $i_m$ are mapped to the initial wave function as $\ket{\varphi^{i_m}}=U_{Cm} \ket{{i_m}} \ket{0}^{\otimes n-1}$, where $U_{Cm}^{i_m}$ and $U_{Cm}$ are unitary operators with polynomial depth in the $m$-th subsystem. The second case can be regarded as a special example of the first case, since $\ket{\varphi^{i_m}}=U_{Cm} \ket{{i_m}} \ket{0}^{\otimes n-1} = U_{Cm} (X_1)^{i_m} \ket{0}^{\otimes n}$ and we can think of $U_{Cm} (X_1)^{i_m}$ as $U_{Cm}^{i_m}$, where $X_1$ is a Pauli $X$ operator which acts on the first qubit. Note that the first method needs a Hadamard test circuit while $M_m$ and $M_{Am}$ can be efficiently constructed via direct measurements in the second case, as will be described later.

\begin{figure}
 \includegraphics[width=1\columnwidth]{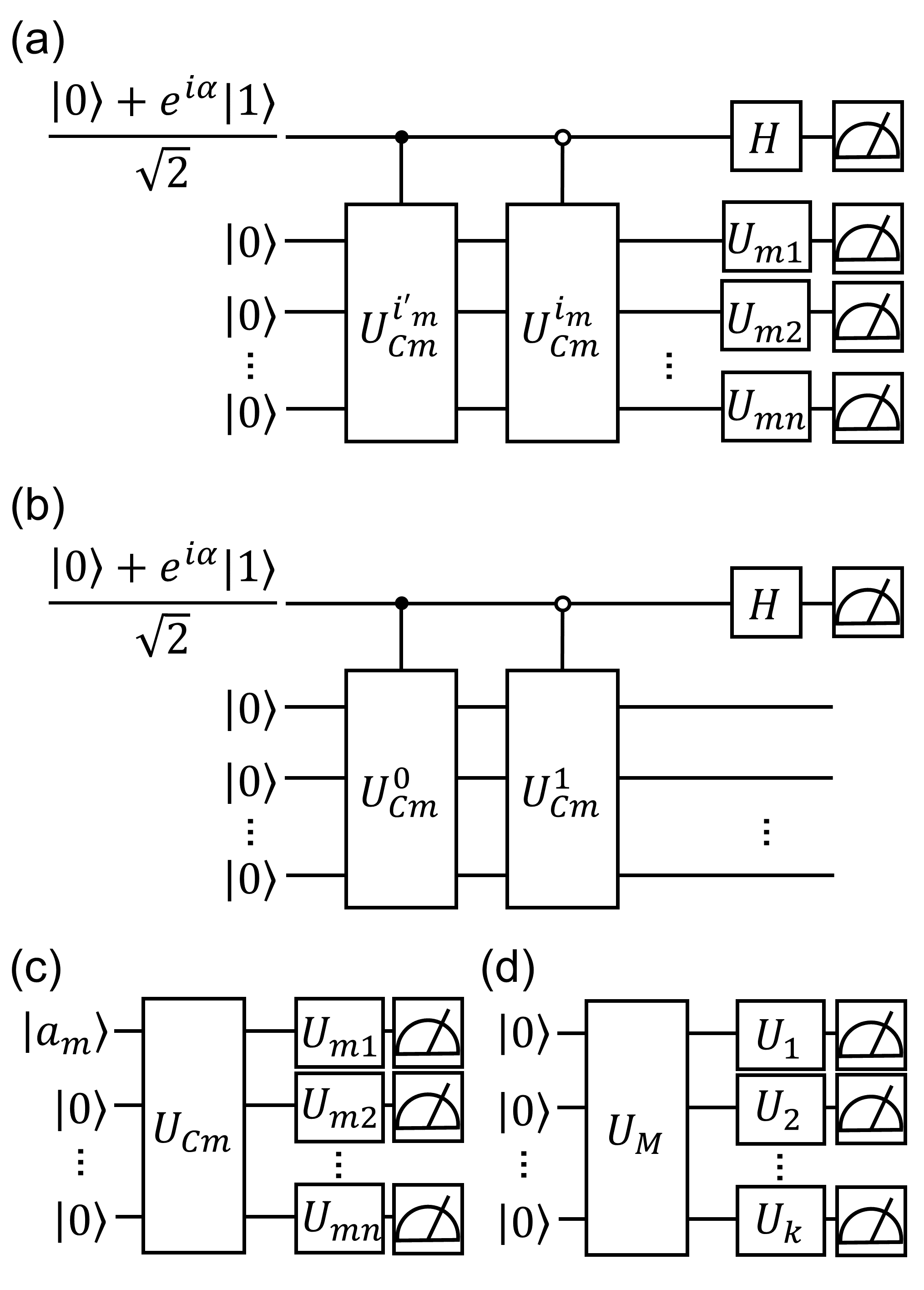}
\caption{Quantum circuits for obtaining $\braket{O}$. In (a) and (b), the topmost line represents an ancilla qubit and the other lines represent system qubits, and in (c) and (d), all lines represent system qubits. Real and imaginary components are obtained by setting $\alpha=0$ and $\alpha=\frac{\pi}{2}$, respectively. A white (black) circle in a controlled gate means that the unitary operation is performed on the target qubits when the control qubit is $\ket{0}$ ($\ket{1}$). (a) A quantum circuit to construct $M_m$ in the case of $\ket{\varphi^{i_m}}=U_{Cm}^{i_m} \ket{0}^{\otimes n}$. (b) A quantum circuit to construct $M_{Am}$ in the case of $\ket{\varphi^{i_m}}=U_{Cm}^{i_m} \ket{0}^{\otimes n}$. (c) A quantum circuit to construct $M_m$ in the case of $\ket{\varphi^{i_m}}=U_{Cm} \ket{{i_m}} \ket{0}^{\otimes n-1}$. $\ket{a_m}$ takes $\ket{0}$, $\ket{1}$, $\ket{+}$ and $\ket{+y}$. (d) A quantum circuit to measure $M_m$ and $M_{Am}$.}
 \label{fig:OverviewOfTTN2}
\end{figure}

First, we consider the construction of $M_m$ in the case of $\ket{\varphi^{i_m}}=U_{Cm}^{i_m} \ket{0}^{\otimes n}$. Since the procedure for measuring the diagonal elements is relatively straightforward, we will focus on the measurement of the non-diagonal elements. Figure~\ref{fig:OverviewOfTTN2}(a) shows a quantum circuit to obtain the matrix element $M_m^{i_m' i_m}$. The procedure for constructing $M_m$ is as follows. First, we prepare initial states. We use a Hadamard test to prepare $\ket{\varphi^{i_m'}}= U_{Cm}^{i_m'} \ket{0}^{\otimes n}$ and $\ket{\varphi^{i_m}} = U_{Cm}^{i_m} \ket{0}^{\otimes n}$. The ancilla qubit is initialized to $\frac{\ket{0}+e^{i\alpha}\ket{1}}{\sqrt{2}}$, where $\alpha$ is the phase. We set $\alpha=0$ ($\alpha=\frac{\pi}{2}$) to obtain the real (imaginary) part of $M_m^{i_m' i_m}$. Since $M_m$ is a Hermitian matrix, only measurements of $\mathrm{Re}(M_m^{0 1})$, and $\mathrm{Im}(M_m^{0 1})$ are required. Then, we measure on a computational basis. We use the fact that $O_{mr}$ is a Hermitian operator and has a spectral decomposition: $O_{mr}=U_{mr}^{\dag} D_{mr} U_{mr}$, where $U_{mr}$ is a unitary matrix and $D_{mr}$ is a diagonal matrix. Also, we assign elements of $D_{mr}$ to the measurement results. More concretely, denoting $D_{mr} = \mathrm{diag}[\lambda^{(mr)}_{j_{r}=0}, \lambda^{(mr)}_{j_{r}=1}]$, the measured value is computed as $\prod_{r=1}^n \lambda^{(mr)}_{j_{r}}$ corresponding to the measured outcome $\vec{j}=(j_1, j_2, \dots ,j_n)$. 

We explain how to construct $M_{Am}$. Figure~\ref{fig:OverviewOfTTN2}(b) shows the circuit to construct $M_{Am}$. In the case of $\ket{\varphi^{i_m}}=U_{Cm}^{i_m} \ket{0}^{\otimes n}$, since the $\ket{\varphi^{i_m}}$ are non-orthogonal to each other, we have $A \neq 1$. Therefore, $M_{Am}$ must be calculated. The circuit in Fig.~\ref{fig:OverviewOfTTN2}(b) is a modified version of that in Fig.~\ref{fig:OverviewOfTTN2}(a), except that system measurements are not necessary; we can construct $M_{Am}$ by using the same construction procedure as for $M_m$.

Next, we assume the case of $\ket{\varphi^{i_m}}=U_{Cm} \ket{{i_m}} \ket{0}^{\otimes n-1}$. In this case, we can obtain all the elements of $M_m$ only from the results of direct measurements. Figure~\ref{fig:OverviewOfTTN2}(c) shows a quantum circuit to construct $M_m$. The initial states are set to $\ket{a_m} \ket{0}^{\otimes n-1}$, where $\ket{a_m}$ takes four states as $\ket{0}$, $\ket{1}$, $\ket{+}$ and $\ket{+y}$. $M_m$ can be obtained using the corresponding measurement results. Refer to Appendix~\ref{sec:The construction of $M_m$ in the case of initial wave function mapping} for the details of the procedure. Also, since $\ket{\varphi^{i_m}}$ is orthogonal, $A = 1$ and $M_{Am}$ does not have to be calculated.

Finally, we show the procedure to obtain $\bra{\psi_{HT}} O \ket{\psi_{HT}}$, which can be implemented by contracting $M_m$ and $M_{A_m}$. Now, denoting $\ket{\psi}=\sum_{\vec{i}} \psi_{\vec{i}} \ket{\vec{i}}=\sum_{i_1, i_2, \dots , i_k} \psi_{i_1, i_2, \dots , i_k} \ket{i_1 i_2 \dots i_k}$, we can rewrite Eq.~(\ref{Eq:tensornetwork2}) as 
\begin{equation}
\begin{aligned}
A^2\braket{O}&= \bra{\psi} \bigotimes_{m=1}^k M_m \ket{\psi} \\
&= \bra{\psi} \bigotimes_{m=1}^k U_m^\dag D_m U_m \ket{\psi},
\label{Eq:expectation value of observable}
\end{aligned} 
\end{equation}
where we have used the fact that $M_m$ is a Hermitian operator and has a spectral decomposition, $M_m=U_m^{\dag} D_m U_m$. Figure~\ref{fig:OverviewOfTTN2}(d) shows a quantum circuit to measure $M_m$. Henceforth, we will assume $\ket{\psi} = U_M \ket{0}^{\otimes n}$, where $U_M$ is a unitary operator with polynomial depth. We can compute $A^2 \braket{O}$ by applying $U_m$, measuring in a computational basis, and assigning elements of $D_m$ to the measurement results. More concretely, denoting $D_m = \mathrm{diag}[\lambda^{(m)}_{i_m=0}, \lambda^{(m)}_{i_m=1}]$, the measured value is computed as $\prod_{m=1}^k \lambda^{(m)}_{i_m}$ corresponding to the measured outcome $\vec{i}=(i_1, i_2, \dots ,i_k)$. In a similar procedure, we can measure $A^2$ by contracting $M_{A_m}$; hence, we can obtain $\braket{O}$. 

\section{Calculation of transition amplitudes and overlaps}
\label{Calculation of transition amplitudes and overlaps}
We describe the measurement of transition amplitudes with the hybrid tensor network approach. The difference from Sec.~\ref{Overview of hybrid tensor network} is that we need to contract a non-Hermitian operator $N_m$.

\begin{figure*}[ht]
 \begin{center}
 \includegraphics[width=2.0\columnwidth]{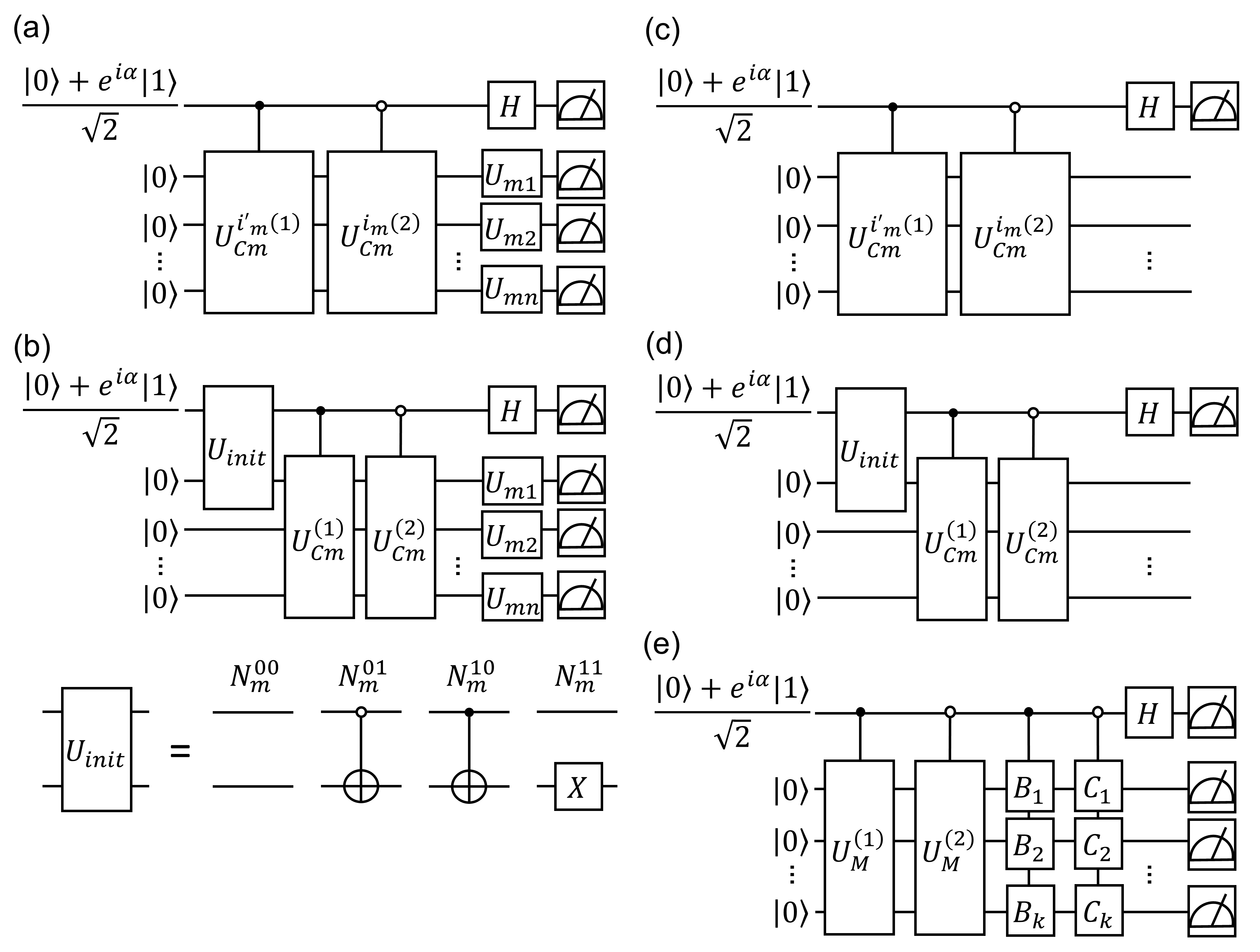}
\caption{Quantum circuits for obtaining transition amplitudes and overlaps. In all the figures, the topmost line represents an ancilla qubit and the other lines represent system qubits. Real and imaginary components are obtained by setting $\alpha=0$ and $\alpha=\frac{\pi}{2}$, respectively. A white (black) circle in a controlled gate means that a unitary operation is performed on the target qubits when the control qubit is $\ket{0}$ ($\ket{1}$). (a) A quantum circuit to construct $N_m$ for calculating transition amplitudes in the case of $\ket{\varphi^{i_m (l)}}=U_{Cm}^{i_m (l)} \ket{0}^{\otimes n}$. (b) A quantum circuit to construct $N_m$ for calculating transition amplitudes in the case of $\ket{\varphi^{i_m (l)}}=U_{Cm}^{(l)} \ket{{i_m}} \ket{0}^{\otimes n-1}$. (c) A quantum circuit to construct $N_m$ for calculating overlaps in the case of $\ket{\varphi^{i_m (l)}}=U_{Cm}^{i_m (l)} \ket{0}^{\otimes n}$. (d) A quantum circuit to construct $N_m$ for calculating overlaps in the case of $\ket{\varphi^{i_m (l)}}=U_{Cm}^{(l)} \ket{{i_m}} \ket{0}^{\otimes n-1}$. (e) A quantum circuit to measure $N_m$ for calculating transition amplitudes or overlaps.}
 \label{fig:TransitionAmplitudeOverlap}
\end{center}
\end{figure*}

\subsection{Contraction of quantum tensors in subsystems}
\label{Contraction of quantum tensors in subsystems}
This section describes construction of $N_m$ in the calculation of $\bra{\psi^{(1)}_{HT}} O \ket{\psi^{(2)}_{HT}}$ in the transition amplitudes and that of $\braket{\psi^{(1)}_{HT}|{\psi^{(2)}_{HT}}}$ in the overlaps,  where $\ket{\psi_{HT}^{(1)}}$ and $\ket{\psi^{(2)}_{HT}}$ are two different states represented by a quantum-quantum tensor network.

To begin with, we will consider the calculation of the transition amplitude because the overlap is a special case of $O=I$ in the transition amplitude, where $I$ is the identity operator. We will comment on the overlap at the end of this section. The transition amplitude $T$ can be defined as
\begin{equation}
\begin{aligned}
T &= \frac{1}{A^{(1)} A^{(2)}} \bra{\psi_{HT}^{(1)}} O \ket{\psi_{HT}^{(2)}} \\
&= \frac{1}{A^{(1)} A^{(2)}} \sum_{\vec{i}'~\vec{i}} \psi_{\vec{i}'}^{(1)*} \psi_{\vec{i}}^{(2)} \prod_{m=1}^k N_m^{i_m' i_m} \\
&= \frac{1}{A^{(1)} A^{(2)}} \bra{\psi^{(1)}} \bigotimes_{m=1}^k N_m \ket{\psi^{(2)}},
\label{Eq:ExpectationNonHermitian1}
\end{aligned}
\end{equation}
where $A^{(l)}$ $(l = 1,2)$ is a normalization constant corresponding to $\ket{\psi_{HT}^{(l)}}$, $\vec{i}= (i_1, i_2, \dots , i_k)$, $N_m$ is a $2 \times 2$ matrix with elements of $N_m^{i_m' i_m} = \bra{\varphi^{i_m' (1)}} O_m \ket{\varphi^{i_m (2)}}$, and $\ket{\psi^{(l)}}=\sum_{\vec{i}} \psi_{\vec{i}}^{(l)} \ket{\vec{i}}=\sum_{i_1, i_2, \dots , i_k} \psi_{i_1, i_2, \dots , i_k}^{(l)} \ket{i_1 i_2 \dots i_k}$. The notation is the same as in Eqs.~(\ref{Eq:tensornetwork2}), (\ref{Eq:tensornetwork3}) and (\ref{Eq:expectation value of observable}), except for the superscript $(l)$, which corresponds to $\ket{\psi_{HT}^{(l)}}$. The reason that $N_m$ is a non-Hermitian matrix comes from the fact that $(N_m^{i_m' i_m})^{*} \neq N_m^{i_m i_m'}$. We will not discuss $A^{(l)}$ in the following since the procedure for calculating $A^{(l)}$ is the same as the one for $A$ in the previous section.

First, let us consider the case of $\ket{\varphi^{i_m (l)}}=U_{Cm}^{i_m (l)} \ket{0}^{\otimes n}$. Figure~\ref{fig:TransitionAmplitudeOverlap}(a) shows the quantum circuit for constructing $N_m$. The flow of constructing $N_m$ is similar to that of $M_m$. However, since $N_m$ is a $2 \times 2$ non-Hermitian matrix with elements of complex numbers, eight types of measurement are required to construct $N_m$. Next, we consider the case of $\ket{\varphi^{i_m (l)}}=U_{Cm}^{(l)} \ket{{i_m}} \ket{0}^{\otimes n-1}$. Figure~\ref{fig:TransitionAmplitudeOverlap}(b) shows a quantum circuit to obtain the matrix element $N_m^{i_m' i_m}$. Specifically, the wave function is initialized by using one of the four types of unitary gates $U_{init}$ in the lower panel in Fig.~\ref{fig:TransitionAmplitudeOverlap}(b) to prepare the initial states $\ket{{i_m'}} \ket{0}^{\otimes n-1}$ and $\ket{{i_m}} \ket{0}^{\otimes n-1}$. The subsequent process is the same as in the case of $\ket{\varphi^{i_m (l)}}=U_{Cm}^{i_m (l)} \ket{0}^{\otimes n}$. 

When calculating the overlap $S$, i.e., the case of $O=I$ in Eq.~(\ref{Eq:ExpectationNonHermitian1}), only the circuits to construct $N_m$ (Figs.~\ref{fig:TransitionAmplitudeOverlap}(a) and (b)) differ from the cases of the transition amplitude. Figures~\ref{fig:TransitionAmplitudeOverlap}(c) and (d) shows the circuits in the cases of $\ket{\varphi^{i_m (l)}}=U_{Cm}^{i_m (l)} \ket{0}^{\otimes n}$ and $\ket{\varphi^{i_m (l)}}=U_{Cm}^{(l)} \ket{{i_m}} \ket{0}^{\otimes n-1}$, which are the same quantum circuits as in Figs.~\ref{fig:TransitionAmplitudeOverlap}(a) and (b) except that the measurements of the system qubits are not involved.
Note that although the calculation of the transition amplitude and the overlap requires the construction of non-Hermitian operators using the ancilla qubit in general, it can be circumvented in the calculation of the square of the overlap by using the destructive SWAP test~\cite{LaRose2019-nl} (Appendix~\ref{sec:square overlaps without ancila} for details).

\subsection{Contraction of non-Hermitian matrices}
\label{Contraction of non-Hermitian matrices}
The next step is to calculate $\bra{\psi^{(1)}} \bigotimes_{m=1}^k N_m \ket{\psi^{(2)}}$ for $2 \times 2$ non-Hermitian matrices $N_m$. We describe two methods of contraction: a Monte-Carlo contraction method and an SVD contraction method. In this section, we describe the SVD contraction method because it is exponentially more efficient than the Monte-Carlo contraction method in terms of the sampling cost. Refer to Appendixes~\ref{sec:Mote-Carlo} and  \ref{sec:comparison} for details of the Monte-Carlo contraction method and the comparison of the two methods, respectively. 

Now let us describe how to perform the SVD contraction method. We perform SVD $N_m=B_m^{\dag}D_m' C_m$ for each $N_m$, that is, $\bigotimes_{m=1}^k N_m = \bigotimes_{m=1}^k B_m^{\dag}D_m' C_m$, where $B_m$ and $C_m$ are unitary matrices and $D_m'$ is a diagonal matrix with non-negative elements. $\bra{\psi^{(1)}} \bigotimes_{m=1}^k N_m \ket{\psi^{(2)}}$ can be described as
\begin{equation}
\begin{aligned}
&\bra{\psi^{(1)}} \bigotimes_{m=1}^k N_m \ket{\psi^{(2)}}\\
& = \bra{\psi^{(1)}} \bigotimes_{m=1}^k B_{m}^{\dag} D'_m C_{m} \ket{\psi^{(2)}}\\
& = (\prod_{m=1}^k \|N_m \|_{op}) \bra{\psi^{(1)}} \bigotimes_{m=1}^k B_{m}^{\dag} d'^{(m)} C_{m} \ket{\psi^{(2)}}\\
& = 2 (\prod_{m=1}^k \|N_m \|_{op}) \cdot \frac{1}{2} \times \big[\mathrm{Re}(\bra{\psi^{(1)}} \bigotimes_{m=1}^k B_{m}^{\dag} d'^{(m)} C_{m} \ket{\psi^{(2)}})\\
&+ i \mathrm{Im}(\bra{\psi^{(1)}} \bigotimes_{m=1}^k B_{m}^{\dag} d'^{(m)} C_{m} \ket{\psi^{(2)}}) \big],
\label{Eq:decomposition of the proposed method}
\end{aligned}
\end{equation}
where $\|A \|_{op}$ is the operator norm of an operator $A$, $d'^{(m)} = D'_m/\|N_m \|_{op}$. We can assume that $d'^{(m)} = \mathrm{diag}[\Tilde{\lambda}^{\prime (m)}_{i_{m}=0}, \Tilde{\lambda}^{\prime (m)}_{i_{m}=1}] = \mathrm{diag}[1, \Tilde{\lambda}^{\prime (m)}_{i_{m}=1}]$, where $\Tilde{\lambda}^{\prime (m)}_{i_{m}=0} \geq \Tilde{\lambda}^{\prime (m)}_{i_{m}=1}$, $\Tilde{\lambda}^{\prime (m)}_{i_{m}=0} = 1$, and $\Tilde{\lambda}^{\prime (m)}_{i_{m}=1}$ takes a value in a range $[0, 1]$, without loss of generality.

Figure~\ref{fig:TransitionAmplitudeOverlap}(e) shows the circuit for measuring $\bra{\psi^{(1)}} \bigotimes_{m=1}^k N_m \ket{\psi^{(2)}}$ by using the SVD contraction method. We can compute $\bra{\psi^{(1)}} \bigotimes_{m=1}^k N_m \ket{\psi^{(2)}}$ as follows. We implement a Hadamard test circuit for $\bigotimes_{m=1}^k B_{m}^{\dag} d'^{(m)} C_{m}$ for obtaining the real or imaginary part of $\bra{\psi^{(1)}} \bigotimes_{m=1}^k N_m \ket{\psi^{(2)}}$ with probability $1/2$ (i.e., 1:1 ratio) by changing the phase of the ancilla qubit. 
We define $\mu_s^{(\mathrm{Re})} = 2 (\prod_{m=1}^k \|N_m \|_{op}) (\prod_{m=1}^k \Tilde{\lambda}^{\prime (m)}_{i_{ms}}) b_s$ and $\mu_s^{(\mathrm{Im})} = 2i (\prod_{m=1}^k \|N_m \|_{op}) (\prod_{m=1}^k \Tilde{\lambda}^{\prime (m)}_{i_{ms}}) b_s$ in the cases of measurements of the real and imaginary parts, respectively, where $i_{ms} \in \{0, 1\}$ and $b_s \in \{-1, 1\}$ are the measurement outcomes of the system and ancilla qubits in the $s$-th measurement, respectively. Then, the sum of the total sample averages of each of $\mu_s^{(\mathrm{Re})}$ and $\mu_s^{(\mathrm{Im})}$ approximates $\bra{\psi^{(1)}} \bigotimes_{m=1}^k N_m \ket{\psi^{(2)}}$.

Letting $ \Bar{x}$ denote the sample average of a random variable $x$, $\mathrm{E}[x]$ denotes the expected value of $x$, and $\Bar{\mu}_s$ = $\Bar{\mu}_s^{(\mathrm{Re})}+\Bar{\mu}_s^{(\mathrm{Im})}$ approximates $\bra{\psi^{(1)}} \bigotimes_{m=1}^k N_m \ket{\psi^{(2)}}$. Denoting the number of measurements as $N_{SVD}$ and assuming $\forall m~\|N_{m}\|_{op}=\|N_{const}\|_{op}$, we have
\begin{align}
&\mathrm{E}[| \Bar{\mu}_s-\bra{\psi^{(1)}} \bigotimes_{m=1}^k N_m \ket{\psi^{(2)}}|]= O\bigg(\frac{(\|N_{const} \|_{op})^k}{\sqrt{N_{SVD}}} \bigg).
\label{Eq:standard deviation of proposed method}
\end{align}
Thus, we have 
\begin{equation}
N_{SVD}= O\bigg(\frac{(\|N_{const}\|_{op})^{2k}}{\varepsilon^2} \bigg)
\label{Eq:the number of qubit in the proposed method}
\end{equation}
for the required accuracy $\varepsilon$. 

We mention that $N_{SVD}$ is expected not to increase exponentially with $k$ if $n$ is large enough. Here, we consider the case where the measured observable $O$ is a product of Pauli operators. This is because in the conventional variational quantum eigensolver (VQE) \cite{Peruzzo2014-kp} scenario, we decompose the Hermitian operator of interest into a linear combination of a polynomial number of products of Pauli operators with the system size. We numerically generate four types of $2^{n} \times 2^{n}$ random unitary matrices, $U^{0 (1)}$, $U^{1 (1)}$, $U^{0 (2)}$, and $U^{1 (2)}$, and a product of random Pauli operators, $O_{rand}$, and create a $2 \times 2$ matrix $N_{const}$ consisting of elements $N^{i' i}_{const} = \bra{0}^{\otimes n} U^{i' (1) \dag} O_{rand} U^{i (2)} \ket{0}^{\otimes n}$, where $i'$ and $i$ take 0 or 1. Then, we evaluate the average values of $\|N_{const}\|_{op}$ obtained using $10,000$ samples of $N_{const}$. As a result, we obtain $\|N_{const}\|_{op} < 1$, including error bars when $n \geq 3$ (See Appendix~\ref{sec:comparison} for details). Therefore, if $n$ is large enough, $N_{SVD} \leq O(1/\varepsilon^2)$ will be valid.

Additionally, we should note that in the case of $\ket{\varphi^{i_m (l)}}=U_{Cm}^{(l)} \ket{{i_m}} \ket{0}^{\otimes n-1}$, $N_{SVD} \leq O(1/\varepsilon^2)$ is strictly satisfied regardless of $n$. In this case, since $N_m$ can be regarded as a submatrix of the unitary matrix $U_{Cm}^{(1)\dag} O_m U_{Cm}^{(2)}$ as $N_m^{i_m' i_m} = \bra{i_m'}\bra{0}^{\otimes n-1} U_{Cm}^{(1)\dag} O_m U_{Cm}^{(2)}\ket{i_m}\ket{0}^{\otimes n-1}$, we have $\|N_m \|_{op} \leq \|O_m \|_{op}$. Thus, because we have assumed $\|O_m \|_{op}=1$ here, we have $N_{SVD} \leq O(1/\varepsilon^2)$ regardless of $n$.

\section{Conclusion}
\label{conclusion}
We proposed a method to calculate transition amplitudes using a hybrid tensor network. When the hybrid tensor network approach is naively applied to the transition amplitude calculation, the contracted operators become non-Hermitian, and the number of terms to be measured increases exponentially. Our method obtains the expectation value without increasing the number of terms exponentially by using singular value decomposition and a Hadamard test. Our theory can be easily generalized to cases with a mixture of classical and quantum tensors called quantum-classical and classical-quantum tree-tensor networks, and those with deeper tree structures. Moreover, we can easily extend the scenario to the case where the measured operator $O$ is a tensor product of non-Hermitian operators by using the SVD contraction method. This study significantly expands the applicability of the hybrid tensor network.

Future work includes the application of our method to algorithms related to hybrid tensor networks. For example, Deep VQE, \cite{Fujii2020-xc,Mizuta2021-pw}, which is a large-scale computational algorithm for NISQ devices based on the divide and conquer method, can be treated in the framework of hybrid tensor networks in theory. By applying the proposed method to such algorithms, we can extend the range of applications to various large-scale quantum algorithms.

\section{Acknowledgments}
This work is supported by PRESTO, JST, Grants No.\,JPMJPR1916 and No.\,JPMJPR2114; ERATO, JST, Grant No.\,JPMJER1601; CREST, JST, Grant No.\,JPMJCR1771; Moonshot R\&D, JST, Grant No.\,JPMJMS2061; MEXT Q-LEAP Grant No.\,JPMXS0120319794 and JPMXS0118068682. S.E. acknowledges useful discussions with Jinzhao Sun and Xiao Yuan.

\bibliographystyle{apsrev4-1}
\bibliography{bib}

\appendix

\section{Construction of $M_m$ in the case where indices $i_m$ are mapped to initial wave functions}
\label{sec:The construction of $M_m$ in the case of initial wave function mapping}
We explain the procedure of constructing $M_m$ in the case of $\ket{\varphi^{i_m}}=U_{Cm} \ket{{i_m}} \ket{0}^{\otimes n-1}$ by using the circuit in Fig.~\ref{fig:OverviewOfTTN2}(c). The diagonal elements, $M_m^{0 0}$ and $M_m^{1 1}$, can be easily obtained by measuring the expectation values of $O_m$ for $\ket{\varphi^{i_m = 0}}$ and $\ket{\varphi^{i_m = 1}}$ because $M_m^{0 0} = \bra{\varphi^{i_m = 0}} O_m \ket{\varphi^{i_m = 0}}$ and $M_m^{1 1} = \bra{\varphi^{i_m = 1}} O_m \ket{\varphi^{i_m = 1}}$, respectively. We can also obtain non-diagonal elements by combining four types of measurement results. By setting $\ket{+^{(m)}}= (\ket{\varphi^{i_m = 0}} + \ket{\varphi^{i_m = 1}})/\sqrt{2}$ and $\ket{+y^{(m)}}= (\ket{\varphi^{i_m = 0}} + i\ket{\varphi^{i_m = 1}})/\sqrt{2}$, we have
\begin{equation}
\begin{aligned}
\bra{+^{(m)}} O_m \ket{+^{(m)}} &= \frac{1}{2}(M_m^{0 0}+M_m^{0 1}+M_m^{1 0}+M_m^{1 1})\\
\label{Eq:expectation value of +}
\end{aligned} 
\end{equation}
and
\begin{equation}
\begin{aligned}
\bra{+y^{(m)}} O_m \ket{+y^{(m)}} &= \frac{1}{2}(M_m^{0 0}+i M_m^{0 1}-i M_m^{1 0}+M_m^{1 1}).\\
\label{Eq:expectation value of y+}
\end{aligned} 
\end{equation}
Then, we can obtain the non-diagonal elements by
\begin{equation}
\begin{aligned}
M_m^{0 1} &= \frac{i-1}{2}\bra{\varphi^{i_m = 0}} O_m \ket{\varphi^{i_m = 0}}\\
 &+ \frac{i-1}{2}\bra{\varphi^{i_m = 1}} O_m \ket{\varphi^{i_m = 1}}\\
 &+\bra{+^{(m)}} O_m \ket{+^{(m)}}\\
 &-i\bra{+y^{(m)}} O_m \ket{+y^{(m)}}
\label{Eq:expectation value of obserbable 1}
\end{aligned} 
\end{equation}
and $M_m^{10}=M_m^{01 *}$.

\section{Calculation of the square of overlaps without the ancilla qubit}
\label{sec:square overlaps without ancila}
In this section, we show the procedure of the calculation for the square of overlaps $|\braket{\psi^{(1)}_{HT}|{\psi^{(2)}_{HT}}}|^2$ by using the destructive SWAP test without using the ancilla qubit.
The main point of the procedure is that we regard $|\braket{\psi^{(1)}_{HT}|{\psi^{(2)}_{HT}}}|^2$ as the expectation value of the SWAP operator for the $2nk$ state $\ket{\Tilde{\psi}_{HT}}= \ket{\psi^{(1)}_{HT}}\ket{\psi^{(2)}_{HT}}$,  where $\ket{\psi_{HT}^{(1)}}$ and $\ket{\psi^{(2)}_{HT}}$ are two different states represented by a quantum-quantum tensor network.
Since the SWAP operator is an observable, we can calculate $|\braket{\psi^{(1)}_{HT}|{\psi^{(2)}_{HT}}}|^2$  by following almost the same procedure as Sec.~\ref{Overview of hybrid tensor network} without the ancilla qubit. 

Specifically, letting the SWAP operator $SWAP=\bigotimes_{m=1}^k SWAP_m$ and $SWAP_m=\bigotimes_{r=1}^n SWAP_{mr}$ and substituting $\ket{\Tilde{\psi}_{HT}}$ for $\ket{\psi_{HT}}$ and $SWAP$ for $O$ in Eq.~(\ref{Eq:tensornetwork2}), the square of overlaps $|\braket{\psi^{(1)}_{HT}|{\psi^{(2)}_{HT}}}|^2$ can be obtained from the expectation value $\braket{SWAP}$ as
\begin{equation}
\begin{aligned}
&\braket{SWAP}\\
&=\bra{\Tilde{\psi}_{HT}} SWAP \ket{\Tilde{\psi}_{HT}}\\
&=\bra{\psi^{(1)}_{HT}}\bra{\psi^{(2)}_{HT}} SWAP \ket{\psi^{(1)}_{HT}}\ket{\psi^{(2)}_{HT}}\\
&=(\bra{\psi^{(1)}_{HT}}\bra{\psi^{(2)}_{HT}}) (\ket{\psi^{(2)}_{HT}}\ket{\psi^{(1)}_{HT}})\\
&= |\braket{\psi^{(1)}_{HT}|{\psi^{(2)}_{HT}}}|^2, \\
\label{Eq:square overlaps1}
\end{aligned}
\end{equation}
and $\braket{SWAP}$ can be also described as
\begin{equation}
\begin{aligned}
&\braket{SWAP}\\
&=  \sum_{\substack{\vec{i}'_{(1)}~\vec{i}'_{(2)}\\ \vec{i}_{(1)}~\vec{i}_{(2)}}}\psi_{\vec{i}'_{(1)}}^{(1)*} \psi_{\vec{i}'_{(2)}}^{(2)*} \psi_{\vec{i}_{(1)}}^{(1)} \psi_{\vec{i}_{(2)}}^{(2)} \prod_{m=1}^k \Tilde{M}_m^{i_{(1)m}' i_{(2)m}' i_{(1)m} i_{(2)m}} \\
&=\bra{\psi^{(1)}}\bra{\psi^{(2)}} \bigotimes_{m=1}^k \Tilde{M}_m \ket{\psi^{(1)}}\ket{\psi^{(2)}}
,
\label{Eq:square overlaps2}
\end{aligned}
\end{equation}
where $\vec{i}_{(l)}= (i_{(l)1}, i_{(l)2}, \dots , i_{(l)k})$ ($l=1,2$) with $i_{(l)m}$ taking either $0$ or $1$, and $\Tilde{M}_m$ is a $4 \times 4$ matrix with elements of 
\begin{equation}
\begin{aligned}
&\Tilde{M}_m^{i_{(1)m}' i_{(2)m}' i_{(1)m} i_{(2)m}} \\
&=\bra{\varphi^{i_{(1)m}' (1)}}\bra{\varphi^{i_{(2)m}' (2)}} SWAP_m \ket{\varphi^{i_{(1)m} (1)}}\ket{\varphi^{i_{(2)m} (2)}}\\
&=\bra{\varphi^{i_{(1)m}' (1)}}\bra{\varphi^{i_{(2)m}' (2)}} \bigotimes_{r=1}^n SWAP_{mr} \ket{\varphi^{i_{(1)m} (1)}}\ket{\varphi^{i_{(2)m} (2)}}.
\end{aligned}
\end{equation}
Here, $\Tilde{M}_m$ is an observable on $m$-th qubits of $\ket{\psi^{(1)}}$ and $\ket{\psi^{(2)}}$, and $SWAP_{mr}$ is an observable on $r$-th qubits of $\ket{\varphi^{i_{(1)m} (1)}}$ and $\ket{\varphi^{i_{(2)m} (2)}}$.
We consider only the case of $\ket{\varphi^{i_{(l)m} (l)}}=U_{Cm}^{(l)} \ket{{i_{(l)m}}} \ket{0}^{\otimes n-1}$, and thus  we omit the normalization constant.

\begin{figure}[t]
 \includegraphics[width=1\columnwidth]{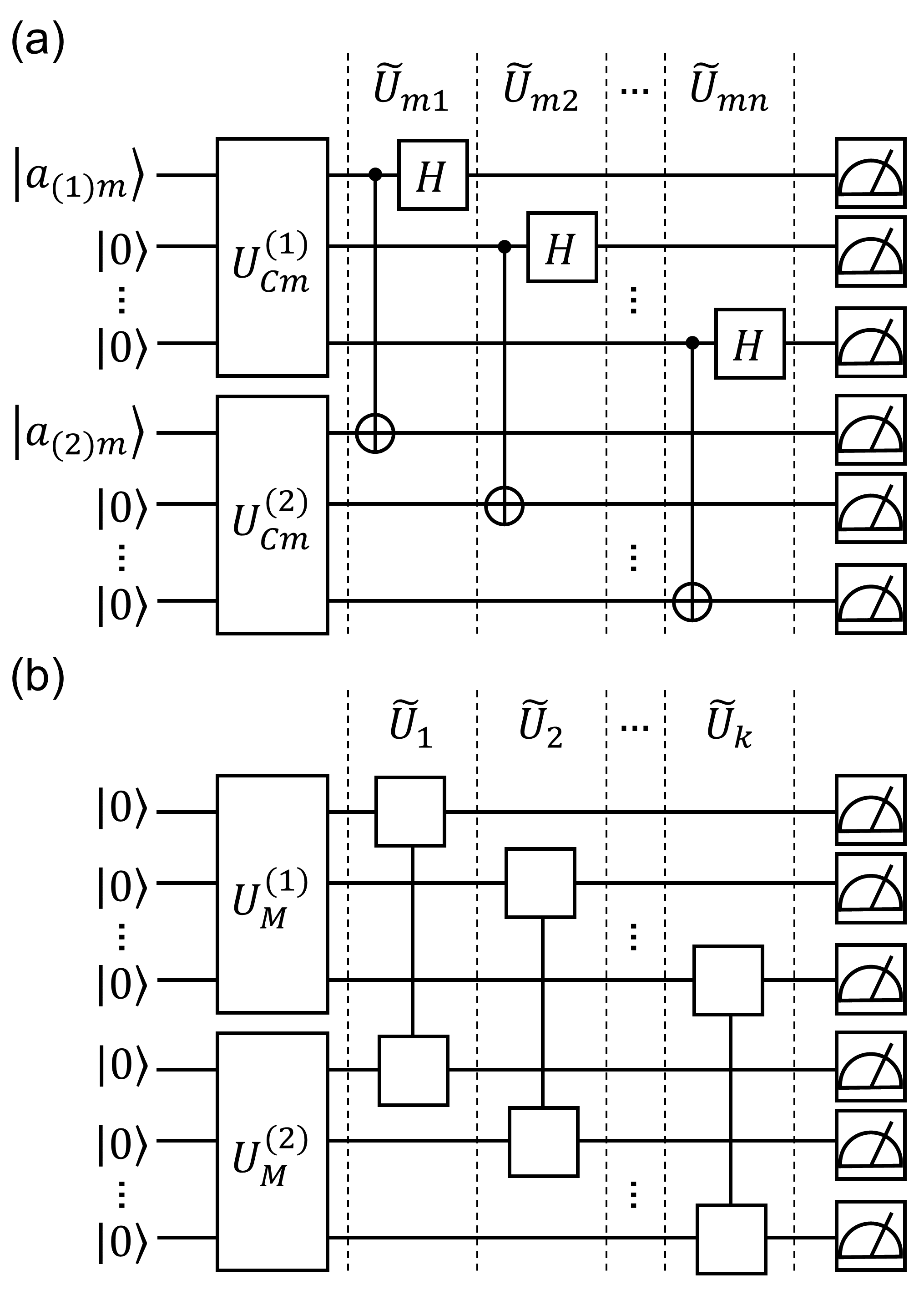}
\caption{
Quantum circuits for obtaining  $|\braket{\psi^{(1)}_{HT}|{\psi^{(2)}_{HT}}}|^2$. The operation in the area enclosed by the dashed line (labeled $\Tilde{U}_{mr}$ in (a) and $\Tilde{U}_m$ in (b)) represents the unitary gate for the measurement for a pair of qubits. (a) A quantum circuit to construct $\Tilde{M}_m$. (b) A quantum circuit to measure $\Tilde{M}_m$.}
 \label{fig:SquareOverlaps}
\end{figure}

Figure~\ref{fig:SquareOverlaps}(a) shows a quantum circuit to obtain the matrix element $\Tilde{M}_m^{i_{(1)m}' i_{(2)m}' i_{(1)m} i_{(2)m}}$ using the destructive SWAP test.
The procedure of obtaining the matrix elements is as follows.
Firstly, the initial states are set to $\ket{a_{(1)m}} \ket{0}^{\otimes n-1} \ket{a_{(2)m}} \ket{0}^{\otimes n-1}$, where $\ket{a_{(l)m}}$ takes four states as $\ket{0}$, $\ket{1}$, $\ket{+}$ and $\ket{+y}$.
To construct $\Tilde{M}_m$, $4\times4=16$ types of the states are required in total (Table~\ref{tab:construction tilde M}).
After preparing the states $(U_{Cm}^{(1)} \otimes U_{Cm}^{(2)}) \ket{a_{(1)m}} \ket{0}^{\otimes n-1} \ket{a_{(2)m}} \ket{0}^{\otimes n-1}$,
unitary gates for the measurement $\Tilde{U}_{mr}$ consisting of CNOT gates and Hadamard gates $H$ are applied. Here, $\Tilde{U}_{mr}$ is obtained by the spectral decomposition of $SWAP_{mr}$ as $SWAP_{mr}=\Tilde{U}_{mr}^{\dag} \Tilde{D}_{mr} \Tilde{U}_{mr}$, where $\Tilde{U}_{mr}$ is a unitary matrix and $\Tilde{D}_{mr}$ is a diagonal matrix.
Finally, we assign elements of $\Tilde{D}_{mr}$ to the measurement results from each qubit pair.

\begin{table*}[t]
\begin{center}
\caption{Construction of matrix elements of $\tilde{M}_m$ by using the 16 types of the measurement results for the $SWAP_m$ operator. 
The results $\Tilde{S}_{a_{(1)m}a_{(2)m}}$ are denoted as $\Tilde{S}_{a_{(1)m}a_{(2)m}} = \bra{a_{(1)m}} \bra{0}^{\otimes n-1} \bra{a_{(2)m}} \bra{0}^{\otimes n-1} (U_{Cm}^{(1)\dag} \otimes U_{Cm}^{(2)\dag}) SWAP_m (U_{Cm}^{(1)} \otimes U_{Cm}^{(2)}) \ket{a_{(1)m}} \ket{0}^{\otimes n-1} \ket{a_{(2)m}} \ket{0}^{\otimes n-1}$, where $a_{(l)m}$ takes $0$, $1$, $+$, and $+y$. }
\label{tab:construction tilde M}
\begin{tabular}{c|c}
\hline
$\Tilde{M}_m^{0000}$  & 
\begin{tabular}{c}
$\Tilde{S}_{00}$
\end{tabular}\\ \hline

$\Tilde{M}_m^{0001}$  & 
\begin{tabular}{c}
$\frac{1-i}{2}(-\Tilde{S}_{00}-\Tilde{S}_{01}+(1+i)\Tilde{S}_{0+}+(1-i)\Tilde{S}_{0(+y)}) $
\end{tabular}\\ \hline

$\Tilde{M}_m^{0010}$  & 
\begin{tabular}{c}
$\frac{1-i}{2}(-\Tilde{S}_{00}-\Tilde{S}_{10}+(1+i)\Tilde{S}_{+0}+(1-i)\Tilde{S}_{(+y)0})  $
\end{tabular}\\ \hline

$\Tilde{M}_m^{0011}$  & 
\begin{tabular}{c}
$-\frac{i}{2}(\Tilde{S}_{00}+\Tilde{S}_{01}+(-1-i)\Tilde{S}_{0+}+(-1+i)\Tilde{S}_{0(+y)}+\Tilde{S}_{10}+\Tilde{S}_{11}+(-1-i)\Tilde{S}_{1+}+(-1+i)\Tilde{S}_{1(+y)}$\\$+(-1-i)\Tilde{S}_{+0}+(-1-i)\Tilde{S}_{+1}+2i\Tilde{S}_{++}+2\Tilde{S}_{+(+y)}+(-1+i)\Tilde{S}_{(+y)0}+(-1+i)\Tilde{S}_{(+y)1}+2\Tilde{S}_{(+y)+}-2i\Tilde{S}_{(+y)(+y)})  $
\end{tabular}\\ \hline

$\Tilde{M}_m^{0100}$  & 
\begin{tabular}{c}
$\Tilde{M}_m^{0001 *}$
\end{tabular}\\ \hline

$\Tilde{M}_m^{0101}$  & 
\begin{tabular}{c}
$\Tilde{S}_{01}$
\end{tabular}\\ \hline

$\Tilde{M}_m^{0110}$  & 
\begin{tabular}{c}
$-\frac{i}{2}(i\Tilde{S}_{00}+i\Tilde{S}_{01}+(-1-i)\Tilde{S}_{0+}+(1-i)\Tilde{S}_{0(+y)}+i\Tilde{S}_{10}+i\Tilde{S}_{11}+(-1-i)\Tilde{S}_{1+}+(1-i)\Tilde{S}_{1(+y)}$\\$+(1-i)\Tilde{S}_{+0}+(1-i)\Tilde{S}_{+1}+2i\Tilde{S}_{++}-2\Tilde{S}_{+(+y)}+(-1-i)\Tilde{S}_{(+y)0}+(-1-i)\Tilde{S}_{(+y)1}+2\Tilde{S}_{(+y)+}+2i\Tilde{S}_{(+y)(+y)})  $
\end{tabular}\\ \hline

$\Tilde{M}_m^{0111}$  & 
\begin{tabular}{c}
$\frac{1-i}{2}(-\Tilde{S}_{01}-\Tilde{S}_{11}+(1+i)\Tilde{S}_{+1}+(1-i)\Tilde{S}_{(+y)1}) $ 
\end{tabular}\\ \hline

$\Tilde{M}_m^{1000}$  & 
\begin{tabular}{c}
$\Tilde{M}_m^{0010 *}$
\end{tabular}\\ \hline

$\Tilde{M}_m^{1001}$  & 
\begin{tabular}{c}
$\Tilde{M}_m^{0110 *}$
\end{tabular}\\ \hline

$\Tilde{M}_m^{1010}$  & 
\begin{tabular}{c}
$\Tilde{S}_{10}$
\end{tabular}\\ \hline

$\Tilde{M}_m^{1011}$  & 
\begin{tabular}{c}
$\frac{1-i}{2}(-\Tilde{S}_{10}-\Tilde{S}_{11}+(1+i)\Tilde{S}_{1+}+(1-i)\Tilde{S}_{1(+y)})  $
\end{tabular}\\ \hline

$\Tilde{M}_m^{1100}$  & 
\begin{tabular}{c}
$\Tilde{M}_m^{0011 *}$ 
\end{tabular}\\ \hline

$\Tilde{M}_m^{1101}$  & 
\begin{tabular}{c}
$\Tilde{M}_m^{0111 *}$
\end{tabular}\\ \hline

$\Tilde{M}_m^{1110}$  & 
\begin{tabular}{c}
$\Tilde{M}_m^{1011 *}$
\end{tabular}\\ \hline

$\Tilde{M}_m^{1111}$  & 
\begin{tabular}{c}
$\Tilde{S}_{11}$
\end{tabular}\\ \hline

\end{tabular}
\end{center}

\end{table*}

Figure~\ref{fig:SquareOverlaps}(b) shows a quantum circuit to construct $\Tilde{M}_m$.
We can obtain $\bra{\psi^{(1)}}\bra{\psi^{(2)}} \bigotimes_{m=1}^k \Tilde{M}_m \ket{\psi^{(1)}}\ket{\psi^{(2)}}$ by preparing $\ket{\psi^{(1)}}\ket{\psi^{(2)}}=(U_M^{(1)} \otimes U_M^{(2)}) \ket{0}^{\otimes k}  \ket{0}^{\otimes k}$
, applying unitary gates for the measurement $\Tilde{U}_m$, and assigning elements of $\Tilde{D}_m$ to the measurement results from each qubit pair, where $\Tilde{M}_m = \Tilde{U}_m^{\dag} \Tilde{D}_m \Tilde{U}_m $, $\Tilde{U}_m$ is a unitary matrix, and $\Tilde{D}_m$ is a diagonal matrix.

Since the destructive SWAP test circumvents the use of ancilla qubits and the effects of additional noise, it should be used in algorithms such as the variational quantum deflation for evaluating excited states~\cite{Higgott2019-az} and variational quantum state diagonalization~\cite{LaRose2019-nl}.

\section{Monte-Carlo contraction method}
\label{sec:Mote-Carlo}
In this section, we introduce a Monte-Carlo contraction method and discuss the sampling cost. We decompose $\bigotimes_{m=1}^k N_m = \bigotimes_{m=1}^k (h_{Im} I_m+h_{Xm} X_m + h_{Ym} Y_m + h_{Zm} Z_m )$, where $I_m$ is an identity operator, $X_m$, $Y_m$ and $Z_m$ are Pauli operators which act on the $m$-th qubit, and $h_\alpha (\alpha \in \{ I_m, X_m, Y_m, Z_m \})$ are the corresponding coefficients. If we expand the last expression, it has an exponentially increasing number of terms with $k$. To circumvent this problem, a Monte-Carlo implementation can be used to calculate $\bra{\psi^{(1)}} \bigotimes_{m=1}^k N_m \ket{\psi^{(2)}}$. From now on, for notational simplicity, we will denote $\sigma^{(m)}_0= I_m$, $\sigma^{(m)}_1=X_m$, $\sigma^{(m)}_2 = Y_m$, $\sigma^{(m)}_3=Z_m$, $h^{(m)}_0 = h_{Im}$, $h^{(m)}_1 = h_{Xm}$, $h^{(m)}_2 = h_{Ym}$, and $h^{(m)}_3 = h_{Zm}$. Introducing $\gamma^{(m)}=\sum_{k=0}^3 |h^{(m)}_k|$, $p^{(m)}_k= |h^{(m)}_k|/\gamma^{(m)}$, $\phi^{(m)}_{i_m} \in \mathbb{R}$, and $e^{i \phi^{(m)}_{i_m}}=h^{(m)}_{i_m}/ |h^{(m)}_{i_m}|$, we have $\sum_{i_m} p^{(m)}_{i_m}=1$ and 
\begin{equation}
\begin{aligned}
&\bra{\psi^{(1)}} \bigotimes_{m=1}^k N_m \ket{\psi^{(2)}} \\
&= 2 (\prod_{m=1}^k \gamma^{(m)}) \sum_{i_1, i_2, ...,i_k} (\prod_{m=1}^k p^{(m)}_{i_m}) \cdot \frac{1}{2} \times e^{i \sum_{m=1}^k \phi^{(m)}_{i_m}} \\
&\times \big[\mathrm{Re}(\bra{\psi^{(1)}} \bigotimes_{m=1}^k \sigma_{i_m}^{(m)} \ket{\psi^{(2)}}) + i \mathrm{Im}(\bra{\psi^{(1)}} \bigotimes_{m=1}^k \sigma_{i_m}^{(m)} \ket{\psi^{(2)}}) \big].
\end{aligned}
\end{equation}
Therefore, we can compute $\bra{\psi^{(1)}} \bigotimes_{m=1}^k N_m \ket{\psi^{(2)}}$ as follows. We generate $\bigotimes_{m=1}^k \sigma_{i_m}^{(m)}$ with probability $\prod_{m=1}^k p^{(m)}_{i_m}$ and implement a Hadamard test circuit for obtaining the real or imaginary part of $\bra{\psi^{(1)}} \bigotimes_{m=1}^k N_m \ket{\psi^{(2)}}$ with probability $1/2$ (i.e., 1:1 ratio) by changing the phase of the ancilla qubit. 
We define $\mu_s'^{(\mathrm{Re})}= 2 (\prod_{m=1}^k \gamma^{(m)}) e^{i \sum_{m=1}^k \phi^{(m)}_{i_m}} b_s'$ and $\mu_s'^{(\mathrm{Im})}= 2i (\prod_{m=1}^k \gamma^{(m)}) e^{i \sum_{m=1}^k \phi^{(m)}_{i_m}} b_s'$ in the cases of measurements of the real and imaginary parts, respectively, where $b_s' \in \{-1, 1\}$ is the measurement outcomes of the ancilla qubit in the $s$-th measurement.
Then, the sum of the total sample averages of each of  $\mu_s'^{(\mathrm{Re})}$ and $\mu_s'^{(\mathrm{Im})}$ approximates $\bra{\psi^{(1)}} \bigotimes_{m=1}^k N_m \ket{\psi^{(2)}}$.

Below, $ \Bar{x}$ denotes the sample average of a random variable $x$, $\mathrm{E}[x]$ denotes the expected value of $x$, and $\Bar{\mu}_s'=\Bar{\mu}_s'^{(\mathrm{Re})}+\Bar{\mu}_s'^{(\mathrm{Im})}$ approximates $\bra{\psi^{(1)}} \bigotimes_{m=1}^k N_m \ket{\psi^{(2)}}$. Denoting the number of measurements as $N_{MC}$ and assuming $\forall m~\gamma^{(m)}=\gamma$, we have
\begin{equation}
\begin{aligned}
\mathrm{E}[| \Bar{\mu}_s'-\bra{\psi^{(1)}} \bigotimes_{m=1}^k N_m \ket{\psi^{(2)}}|]= O(\gamma^k/\sqrt{N_{MC}}).
\label{Eq:standard deviation of monte carlo}
\end{aligned}
\end{equation}
Thus, we need 
\begin{equation}
N_{MC}= O(\gamma^{2k}/\varepsilon^2) 
\label{Eq:the number of sample in mote carlo}
\end{equation}
for the required accuracy $\varepsilon$. 

\section{Comparison of Monte-Carlo contraction method and SVD contraction method}
\label{sec:comparison}
Here, we compare $N_{MC}$ with $N_{SVD}$. Since $N_m=\sum_{i_m} h_{i_m} \sigma_{i_m}^{(m)}$, we have
\begin{align}
\|N_m \|_{op} \leq \sum_{i_m}|h_{i_m}| \|\sigma_{i_m}^{(m)} \|_{op}= \gamma^{(m)},
\label{Eq: Nop < gamma}
\end{align}
where we have used $\|\sigma_{i_m}^{(m)} \|_{op}=1$.

Equations~(\ref{Eq:the number of qubit in the proposed method}), (\ref{Eq:the number of sample in mote carlo}), and (\ref{Eq: Nop < gamma}) indicate $\gamma / \|N_{const} \|_{op} \geq 1$ and 
\begin{equation}
 \begin{aligned}
 \frac{N_{MC}}{N_{SVD}}=O((\gamma / \|N_{const} \|_{op})^{2k}).
 \label{Eq: gamma/Nop}
 \end{aligned}
\end{equation}
Thus, the SVD contraction method is exponentially more efficient than the Monte-Carlo contraction method in terms of the sampling cost.

\begin{figure}[H]
 \vspace{0.5cm}
 \includegraphics[width=1\columnwidth]{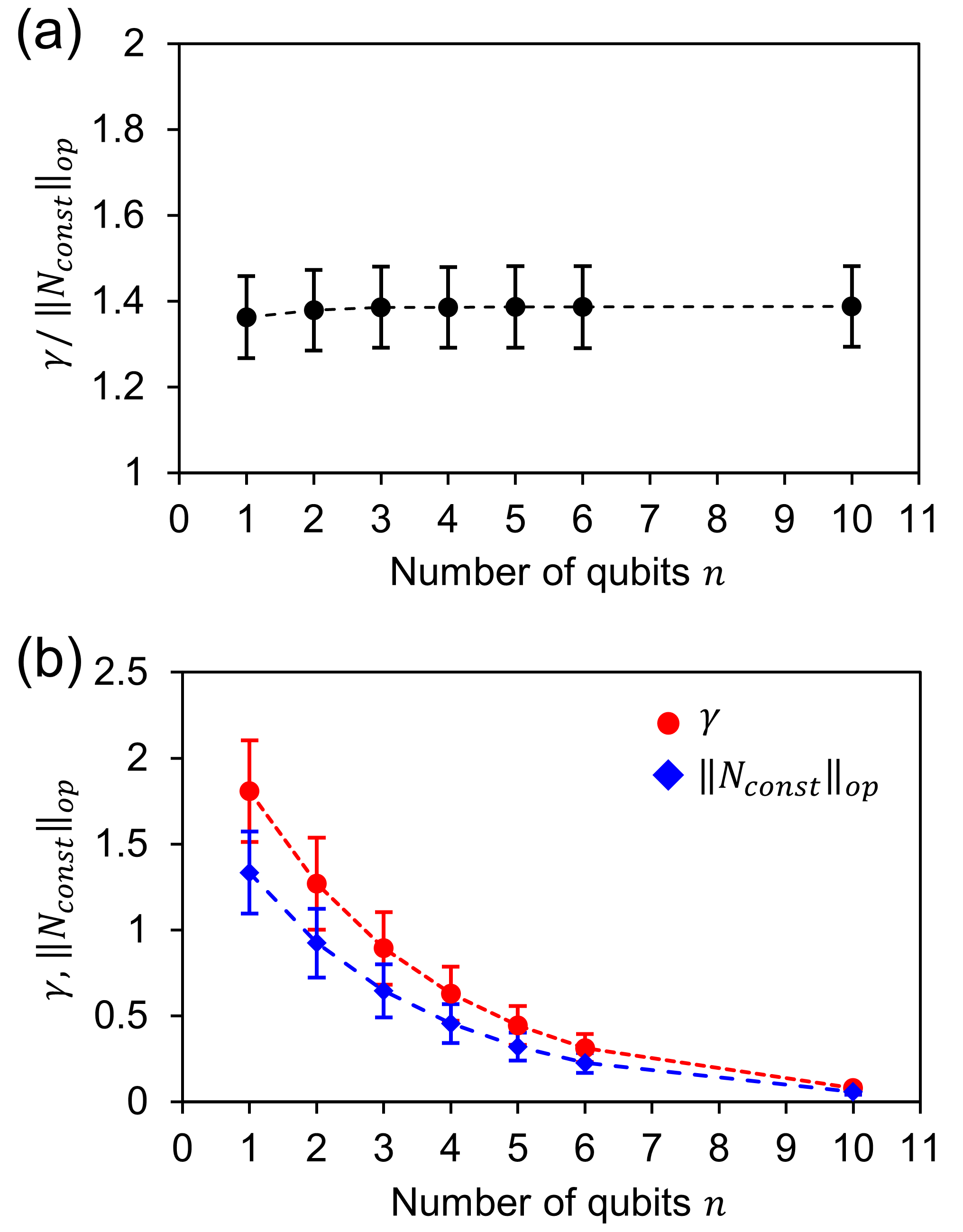}
\caption{Average values of $\gamma$ / $\|N_{const} \|_{op}$, $\gamma$, and $\|N_{const} \|_{op}$ depending on the number of qubits $n$ (10,000 samples). Each point and error bar represent the average value and standard deviation of the samples, respectively. (a) Average value of $\gamma$ / $\|N_{const} \|_{op}$. (b) Average values of $\gamma$ (circule, red) and $\|N_{const} \|_{op}$ (diamond, blue).}
 \label{fig:GammaNopRatio_1col}
\end{figure}

We present numerical calculations for $\gamma / \|N_{const} \|_{op}$ in order to check the superiority of the SVD contraction method over the Monte-Carlo contraction method. We obtain $10,000$ samples of $N_{const}$ and $\|N_{const} \|_{op}$ by the procedure in Sec.~\ref{Contraction of non-Hermitian matrices} and $\gamma$ from the Pauli decomposition of $N_{const}$. Fig.~\ref{fig:GammaNopRatio_1col}(a) shows the average of the ratio $\gamma / \|N_{const} \|_{op}$; we found that the average value is about $1.4$ for any $n$. Therefore, from Eq.~(\ref{Eq: gamma/Nop}) and the result in Fig.~\ref{fig:GammaNopRatio_1col}(a), we can conclude that the SVD contraction method is expected to be $O((1.4)^{2k})$ times faster on average than the Monte-Carlo contraction method.

We also numerically evaluate the number of measurements for the two methods. The averages of $\gamma $ and $\|N_{const} \|_{op}$ are shown in Fig.~\ref{fig:GammaNopRatio_1col}(b). $\gamma$ and $\|N_{const} \|_{op}$ in $n \geq 4$ and $n \geq 3$, respectively, can be considered to be less than 1, including the standard deviation. Thus, we expect $N_{MC} \leq O(1/\varepsilon^2)$ and $N_{SVD} \leq O(1/\varepsilon^2)$; that is, $N_{MC}$ and $N_{SVD}$ are not expected to increase exponentially with $k$, if $n$ is large enough.

\end{document}